\documentclass[amsmath,amssymb,prX,superscriptaddress,twocolumn]{revtex4-1}

\usepackage{graphicx}
\usepackage{dcolumn}
\usepackage[colorlinks=false]{hyperref}
\usepackage[sort&compress]{natbib}

\begin{document}
\title{THz lattice vibrations for active plasmonics with light: \\ Ultrafast optical response in gold/telluride hybrid plasmonic crystals}

\author{Lars~E.~Kreilkamp}\email{lars.kreilkamp@tu-dortmund.de}
\affiliation{Experimentelle Physik 2, Technische Universit\"at Dortmund, D-44221 Dortmund, Germany}
\author{Ilya~A.~Akimov}
\affiliation{Experimentelle Physik 2, Technische Universit\"at Dortmund, D-44221 Dortmund, Germany}
\affiliation{Ioffe Institute, Russian Academy of Sciences, 194021 St. Petersburg, Russia}	
\author{Vladimir~I.~Belotelov}
\affiliation{Lomonosov Moscow State University, 119991 Moscow, Russia}
\affiliation{Russian Quantum Center, 143025 Skolkovo, Moscow Region, Russia}
\author{Boris~A.~Glavin}
\affiliation{Lashkaryov Institute of Semiconductor Physics, 03028 Kyiv, Ukraine }
\author{Leonid~Litvin}
\author{Axel~Rudzinski}
\author{Michael~Kahl}
\author{Ralf~Jede}
\affiliation{Raith GmbH, Konrad-Adenauer-Allee 8, 44263 Dortmund, Germany}
\author{Maciej~Wiater}
\author{Tomasz~Wojtowicz}
\author{Grzegorz~Karczewski}
\affiliation{Institute of Physics, Polish Academy of Sciences, PL-02668 Warsaw, Poland }
\author{Dmitri~R.~Yakovlev}
\author{Manfred~Bayer}
\affiliation{Experimentelle Physik 2, Technische Universit\"at Dortmund, D-44221 Dortmund, Germany}
\affiliation{Ioffe Institute, Russian Academy of Sciences, 194021 St. Petersburg, Russia}
\date{\today}

\begin{abstract}
Excitation of coherent optical phonons in solids provides a pathway for ultrafast modulation of light on a sub-ps timescale. Here, we report on efficient 3.6\,THz modulation of light reflected from hybrid metal/semiconductor plasmonic crystals caused by lattice vibrations in a few nm thick layer of elemental tellurium. We observe that surface plasmon polaritons contribute significantly to photoinduced formation of this thin layer at the interface between a telluride-based II-VI semiconductor, such as (Cd,Mg)Te or (Cd,Mn)Te, and a one-dimensional gold grating.
The change in interface composition is monitored via the excitation and detection of coherent optical tellurium phonons of ${\rm A}_1$ symmetry by femtosecond laser pulses in a pump-probe experiment.
The patterning of a plasmonic grating onto the semiconductor enhances the transient signal which originates from the interface region. This allows monitoring the layer formation and observing the shift of the phonon frequency caused by confinement of the lattice vibrations in the nm-thick segregated layer. Efficient excitation and detection of coherent optical phonons by means of surface plasmon polaritons are evidenced by the dependence of the signal strength on polarization of pump and probe pulses and its spectral distribution.
\end{abstract}



\maketitle

\section{Introduction}
\label{sec1}
Strong localization of light in plasmonic structures is widely used to intensify a large variety of optical phenomena. One of the most prominent examples is surface enhanced Raman scattering where the use of metallic nanoparticles in direct vicinity of the investigated material allows increasing the effective Raman cross section by many orders of magnitude and reaching the signal levels capable of single molecule detection~\cite{Moskovits}. Surface plasmon polaritons (SPPs) at the interface between metals and dielectrics can be efficiently used for enhancement of light-matter interaction as well~\cite{Zayats12}. First, the excitation of a SPP leads to significant electromagnetic energy localization near the metal-dielectric interface. Second, the polariton dispersion is very sensitive to variations of the dielectric function in a sub-wavelength thin region close to the interface. Based on these principles, surface plasmon resonant sensors for the detection of chemical and biological species are developed~\cite{Homola}. In addition, one can use an external stimulus such as optical excitation~\cite{Pacifici07, MacDonald, Brongersma, Caspers10, Pohl12}, acoustic waves~\cite{delFatti99, Hodak99, Ruppert10, Bruegemann12} and external magnetic fields~\cite{Armelles, Temnov12, Akimov2012,Belotelov14} in order to influence the dielectric function of metal and/or dielectric and subsequently to control the SPP propagation constant. Therefore, SPPs are well suited for applications in optoelectronic devices where high sensitivities and efficient optical excitation at the nanoscale are essential~\cite{Brongersma}. In this way, plasmonic crystals--periodic plasmonic materials--are mostly promising since their structure allows the excitation of different type of modes including propagating and localized SPPs as well as cavity and hybrid plasmon-waveguide modes~\cite{Christ03,Belotelov12,Lars13}.

The combination of plasmonic structures with semiconductor materials is of great interest for both applied and fundamental science. Semiconductor nanostructures with their excellent optical and electrical properties can be tailored for the targeted applications on a detailed level by nanotechnology \cite{nanostructures-book}. In their turn, plasmonic structures can be used to increase the strength of the dynamical optical response from semiconductor nanostructures and to uncover new processes which are typically too weak to be observed in conventional systems. Optical studies in hybrid plasmonic-semiconductor structures demonstrated SPP assisted photoinjection of hot carriers from the metal into the semiconductor~\cite{Brongersma,Hallas11}. Some progress in coupling the fundamental optical excitations in semiconductors with SPPs has been achieved, e.g. modifications of the exciton energy spectrum~\cite{Vasa08, Bellessa13} and recombination dynamics~\cite{Okamoto09, Andersen11} in direct band gap III-V compounds were reported. However, ultrafast studies with optical pulses, which deserve special attention in connection with active plasmonics, have been limited to a few works where the propagation constant of SPPs is controlled due to the modulation of the dielectric function by photoexcited carriers in a semiconductor layer~\cite{Pacifici07,Caspers10}.

In this paper, we demonstrate that SPPs can be successfully used for efficient excitation and detection of high frequency coherent optical phonons in semiconductors at the nanoscale. And vice versa, lattice vibrations allow performing ultrafast THz control of SPP optical resonances. In case of the considered plasmonic crystals this is accomplished by THz modulation of the SPP resonance quality factor. The results are based on ultrafast time-resolved pump-probe studies in hybrid structures which comprise an epitaxially grown (Cd,Mg)Te or (Cd,Mn)Te layer capped with a one-dimensional plasmonic grating. The experiments are performed in the transparency region, i.e. the photon energy $\hbar\omega_0\approx 1.5\,$eV is significantly lower than the energy gap in the semiconductor layer $E_{\rm g} > 1.7\,$eV. The patterning of a noble metal grating on top of a semiconductor allows for the excitation of propagating SPPs at the metal-semiconductor interface, which gives rise to the observation of two main features. First, we detect the formation of a tellurium (Te) layer at the metal-semiconductor interface due to optically induced segregation. Its formation is evidenced by coherent oscillations of the symmetric ${\rm A}_1$ breathing mode in differential reflectivity transients with a frequency of about $3.6\,$THz. The buildup of this layer can be monitored over time by analyzing the change in oscillation amplitude and frequency during prolonged exposure to the pump beam. In contrast to the results obtained on the plasmonic sample, the oscillatory signal is barely visible when no gold grating is patterned onto the semiconductor. Second, by performing polarization- and spectrally-resolved measurements for different angles of light incidence, we find that SPPs lead to an enhancement of both, excitation of coherent phonons and their detection in the differential reflectivity. The high sensitivity of the plasmonic structure to vibrations of the crystal lattice allows measuring signals from a 1\,nm thin layer of Te, where confinement of optical phonons is manifested in a shift of the ${\rm A}_1$ phonon frequency to higher values.

\section{Experimental approach}
\label{sec2}

The semiconductor structures were grown using molecular-beam epitaxy on semi-insulating GaAs substrates having either a crystal orientation of (001) or (111) along the growth direction (axis $z$). The structures contain a several $\mu$m thick CdTe-buffer layer which is introduced in order to reduce the mismatch in lattice constant between the substrate and the active $\mu$m thick layer of ternary (Cd,Mg)Te or (Cd,Mn)Te alloy. Pump-probe data give similar results on all of the samples. Further on, we concentrate on a (001) oriented sample with $4\,{\rm \mu m}$ thick CdTe-buffer layer followed by $1.1\,{\rm \mu m}$ thick Cd$_{0.86}$Mg$_{0.14}$Te layer (\#100112A). Photoluminescence and reflectivity spectra indicate an energy gap of about $E_{\rm g} = 1.74\,$eV (713\,nm) at room temperature $T=300$\,K and $E_{\rm g} = 1.82\,$eV (681\,nm) at $T=10$\,K, which is in agreement with 14\% Mg content~\cite{EunsoonOh1993}.

\begin{figure}
	\centering
	\includegraphics[width=\linewidth]{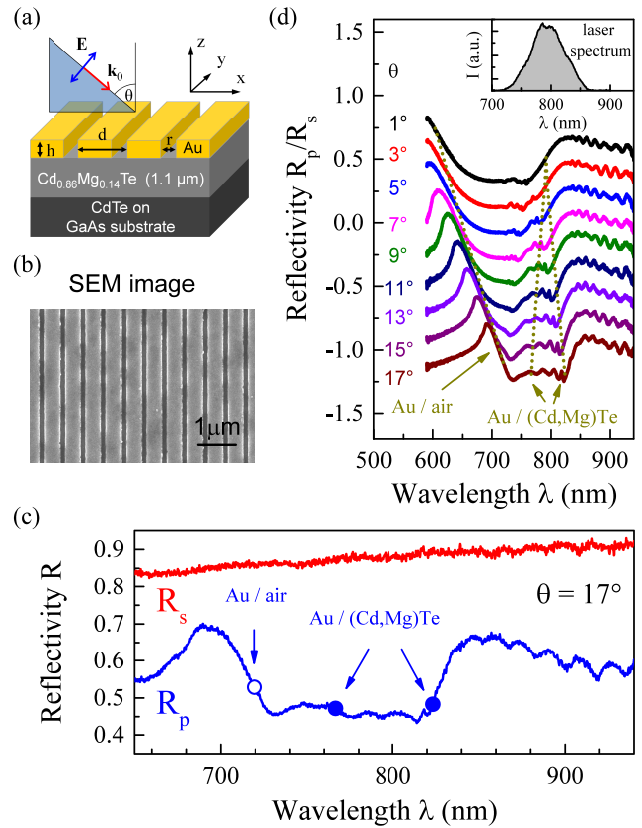}
	\caption{(a) Schema of the investigated samples and the geometry of experiment used for reflectivity measurements with spectrally broad white light source. (b) SEM image of the surface of the plasmonic grating. (c) Reflectivity spectra taken on the plasmonic crystal in $p$- and $s$- polarizations and normalized to reflectivity from a flat gold surface. Angle of incidence is $17^\circ$. Symbols represent the calculated resonance positions of the excited SPPs at the gold-air interface (open circle) and at the gold-semiconductor interface (filled circles). (d) Reflectivity spectra taken under white light illumination on the plasmonic grating for varying incidence angles $\theta$. Spectra measured in the plasmonic $p$-polarization $R_p(\lambda)$ were normalized to the spectra in the non-plasmonic $s$-polarization $R_s(\lambda)$. Spectra are offset by 0.2 for better visibility. The dotted lines indicate the  calculated positions of the SPP resonances. Inset: spectrum of the Ti:Sa laser.}
	\label{fig1}
\end{figure}

The metallic gratings on top of the semiconductor were prepared by electron-beam lithography (EBL) of a 300\,nm thick poly (methyl methacrylate) (PMMA) resist, gold evaporation and a subsequent lift-off procedure. For patterning the photoresist, the EBL instrument RAITH150$^{\rm TWO}$ was used. A schematic sketch of the investigated structure and its main parameters are shown in Fig.~\ref{fig1}(a). The scanning electron microscope (SEM) image of a gold grating is shown in Fig.~\ref{fig1}(b). The grating parameters are: height of gold $h\approx 70\text{--}100$\,nm, grating period $d = 525$\,nm and slit width $r = 100$\,nm. A part of the sample surface was left uncovered by gold and PMMA, thus leaving areas of bare (Cd,Mg)Te serving as a reference for the optical experiments.

The investigated structures can be considered as plasmonic crystals because the periodically patterned metal film attached to the semiconductor gives rise to interference of SPPs and their coupling to far field radiation. The spectral position of SPP resonances and their dispersion follow from angle-resolved zero-order reflectivity spectra which are shown in Fig.~\ref{fig1}(c) and \ref{fig1}(d). The measurements were performed using a spectrally broad white light source (tungsten lamp) and the plane of incidence was perpendicular to the grating slits [see Fig.~\ref{fig1}(a)]. The polarization configuration shown in Fig.~\ref{fig1}(a) corresponds to the $p$-polarized case, where the electric field $\vec{E}$ of the incoming light wave of wavevector $\vec{k_0}$ is oscillating parallel to the plane of incidence. Fig.~\ref{fig1}(c) shows reflectivity spectra measured for $p$-polarized incident light along with the spectrum measured for $s$-polarized incident light under an incidence angle of $\theta=17^\circ$. The experimental data are normalized to the reflectivity spectrum of a flat gold surface illuminated under the same angle with light of the corresponding polarization. The resonance positions of the SPPs excited at the gold-air (gold-semiconductor) interface in the case of $p$-polarized light are marked as open (filled) dots.

It follows that in $s$-polarization more than 80\% of light is reflected back and the spectrum $R_s(\lambda)$ is flat. Indeed, TM polarized SPPs can be excited only in $p$-polarization. The $p$-polarized spectrum $R_p$ contains several resonances with an asymmetric Fano-like shape. High frequency oscillations originate from Fabry-Perot interferences in the layered structure. Reflectivity spectra $R_p(\lambda)/R_s(\lambda)$ for various incident angles are presented in Fig.~\ref{fig1}(d). Here, $p$-polarized spectra are normalized by featureless $s$-polarized curves. The spectral position of the resonances and their angular dependence correspond well to the calculated frequencies of the SPP modes at the gold-air (around 600--700\,nm) and gold-semiconductor (750--850\,nm) interfaces [see dotted lines in Fig.~\ref{fig1}(d)]. The SPP frequencies were calculated by the scattering-matrix method~\cite{Tikh02}. The spectral dispersion of both the dielectric constant of Cd$_{0.86}$Mg$_{0.14}$Te~\cite{Lugauer94} and gold~\cite{Johnson72} were taken into account. Here, the best accordance of optical spectra and modeling were found for grating parameters $d \approx 525$\,nm, $r \approx 170$\,nm and $h \approx 100$\,nm. The difference in slit width $r$ when compared with the SEM data is caused by possible deviations of the dielectric function in the semiconductor constituent close to the interface and/or a non-rectangular profile of the metal grooves.

The temporal dependence of the reflectivity was studied using an ultrafast pump-probe setup. Here, a Ti:Sa oscillator (Coherent MICRA) generated optical pulses with a central wavelength of 800\,nm, spectral width of about 70\,nm and a repetition rate of 80\,MHz. The laser beam was divided by a silica beam splitter into a pump and a probe beam. The temporal delay between pump and probe pulses was adjusted by a mechanical delay line mounted in the pump beam path. Polarization optics in both beam paths were used to adjust the intensity and the polarization of the two laser beams. The beams were subsequently focused onto the sample using a reflective microscope objective with a magnification factor of 15 comprising 4 sectors through which light could enter and exit the objective. The incidence angles of the laser beams were $\theta \approx 17^\circ$. In order to ensure that the probe beam addresses a homogeneously excited sample area, the spot diameters for pump and probe beam were set to $13\,{\rm \mu m}$ and $10\,{\rm \mu m}$, respectively. The energy density of the pump pulse was varied in the range from 30 to $140\,{\rm \mu J/cm^2}$ and the probe fluence was set to smaller values in the range from 8 to $20\,{\rm \mu J/cm^2}$. The pulse chirp acquired during propagation through the optical elements was compensated by means of a pulse shaper and a compressor using the multiphoton intrapulse interference phase scan procedure~\cite{MIPS}. The pulse duration of the transform limited pulses was about 20\,fs. For low temperature measurements at 10\,K the sample was mounted on the cold finger of a He flow cryostat. The reflected light intensity was detected by one of the diodes of a balanced photodetector while the second diode was exposed with a fraction of the laser beam as a reference to account for intensity fluctuations of the Ti:Sa laser. The transients were obtained by modulating the pump intensity with a mechanical chopper wheel and demodulating the differential photodetector signal by means of a lock-in amplifier. For spectral dependence measurements an additional tunable interference filter was introduced into the probe beam before the detector.

\begin{figure}
	\centering
	\includegraphics[width=\linewidth]{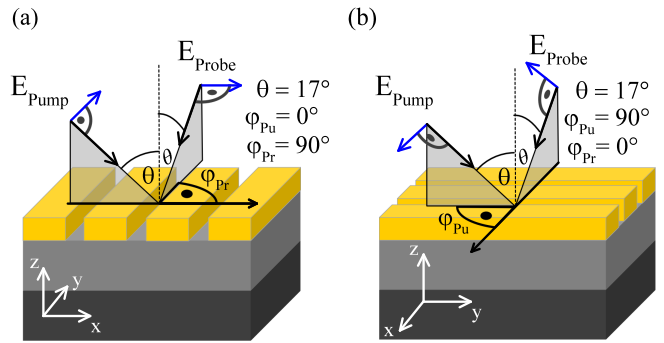}
	\caption{Sketch of the two geometries used for the pump-probe measurements. Polarization configuration is set such that SPPs can be excited by both pump and probe beams. (a) The pump beam is incident perpendicular to the grating slits, while the probe beam is incident parallel to the slits. (b) The pump beam is incident parallel to the grating slits, while the probe beam is incident perpendicular to the slits.}
	\label{fig2}
\end{figure}

The investigated structures support SPPs propagating mainly perpendicular to the grating slits and therefore the mutual orientation of the grating axis $x$ and the plane of incidence for the pump and probe beams plays a crucial role. In order to spatially separate the reflected beams, the pump and probe pulses were sent into the microscope objective through different neighboring sectors. Therefore, the planes of incidence of both beams were orthogonal to each other. In this case, there exist two possible geometries as depicted in Fig.~\ref{fig2}. First, the pump beam is impinging onto the sample perpendicular to the grating slits ($\varphi_{\rm pu}=0^\circ$) while the plane of incidence of the probe beam is parallel to the slits ($\varphi_{\rm pr}=90^\circ$). By rotating the sample by $90^\circ$ the second geometry is realized. Here, the situation is opposite with $\varphi_{\rm pu}=90^\circ$ and $\varphi_{\rm pr}=0^\circ$.

Although the incidence angle $\theta=17^\circ$ is equal for both of the beams, the spectral positions of SPPs that are addressed by pump and probe beams are different. For azimuthal angle $\varphi=0^\circ$ the excitation of SPPs takes place along the $x$ axis with wavevector $k_{\rm SPP} = k_0 \sin\theta \pm mG$, where $G = 2\pi/d$ is the reciprocal vector magnitude of the plasmonic crystal lattice and $m$ is an integer. In this case SPP resonances which are covered by our pulsed laser source are located at 766\,nm and 823\,nm as follows from Fig.~\ref{fig1}(c) and \ref{fig1}(d).

For $\varphi=90^\circ$ SPPs are also excited predominantly along the $x$-axis with $k_{\rm SPP} \approx \pm \sqrt{\left(mG\right)^2+\left(k_0\sin{\theta}\right)^2}\approx\pm mG$ since $\left(k_0\sin\theta\right)^2 \ll m^2G^2$ holds. In this second geometry, the excitation of SPPs is approximately equivalent to the case of normal incidence ($\theta=0^\circ$) and corresponds to a resonance around 792\,nm [see Fig.~\ref{fig1}(d)]. The use of both geometries allowed us to evaluate the role of SPPs by comparing the spectral dependence of the ultrafast response when different SPP resonances are addressed. These results are presented in section~\ref{sec4}. Note that all SPPs addressed in the pump-probe experiment originate from the interface between gold and semiconductor.

\section{SPP assisted Tellurium segregation at the interface}
\label{sec3}

\begin{figure}
	\centering
	\includegraphics[width=\linewidth]{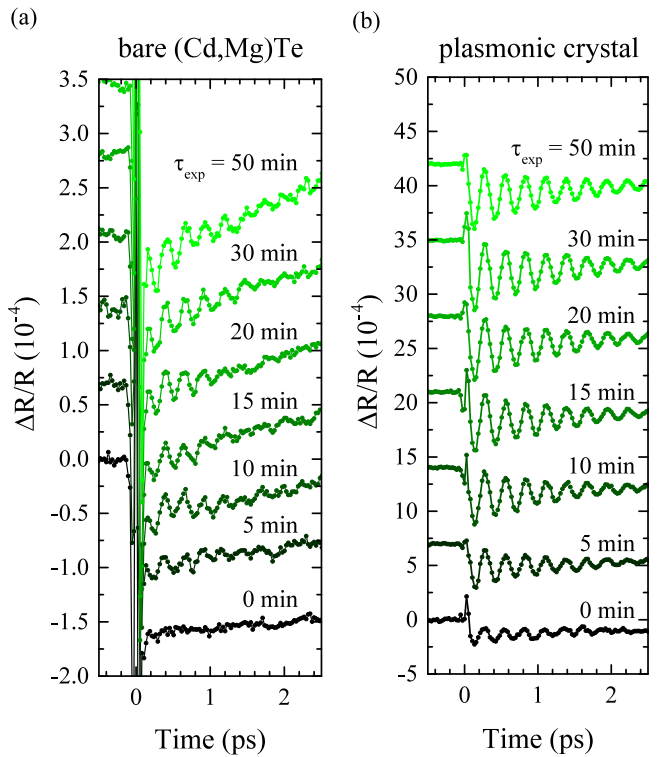}
	\caption{Room temperature differential reflectivity signals after exposure to a pump fluence of $140\,{\rm \mu J/cm^2}$ for varying exposure times $\tau_{\rm exp}$. The transients were taken at a pump fluence of $30\,{\rm \mu J/cm^2}$. The left set of data (a) was acquired on an area of bare (Cd,Mg)Te, while the right set of data (b) was acquired on the plasmonic crystal structure. Data are shifted vertically for clarity.}
	\label{fig3}
\end{figure}

In this section we discuss the ultrafast response in the Au/semiconductor hybrid plasmonic crystals measured in the first configuration as depicted in Fig.~\ref{fig2}(a) ($\varphi_{\rm pu} = 0^\circ$ and $\varphi_{\rm pr} = 90^\circ$). The pump and the probe beams were $p$- and $s$- polarized, respectively, in order to establish efficient interaction with SPPs. A selection of pump induced differential reflectivity transients $\Delta R(t)/R$ measured at room temperature is shown in Fig.~\ref{fig3}. The left set of data (a) presents measurements performed on the bare semiconductor while the right set of data (b) was obtained by illuminating a part of the sample covered with a plasmonic crystal. The measurements were performed with a low pump fluence of $\Phi_{\rm pu} = 30\,{\rm \mu J/cm^2}$. In between measurements, the sample was successively exposed to $\Phi_{\rm pu} = 140\,{\rm \mu J/cm^2}$. The cumulative exposure time $\tau_{\rm exp}$ increases from the lower curves (black) to the upper curves (light green). The impact of the pump beam at the time $t=0$ leads to the appearance of high frequency THz oscillations in the reflectivity transients which decay on a ps-timescale. In addition, a non-oscillatory exponential decay is present in the transients. While the oscillations are only faintly visible for the bare semiconductor, the excitation of the plasmonic crystal leads to an enhancement of the signal by one order of magnitude with the amplitude reaching values of up to $\frac{\Delta R}{R}=3\times 10^{-4}$.

\begin{figure*}
	\centering
	\includegraphics[width=0.9\linewidth]{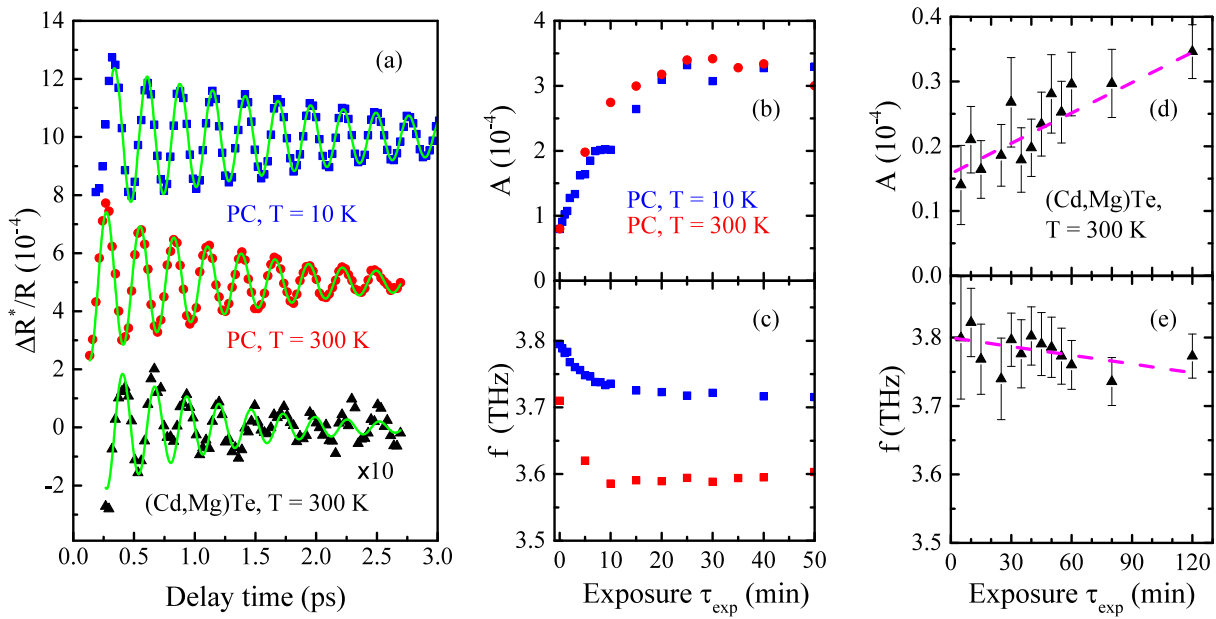}
	\caption{ (a) Oscillatory part of the transient signal $\Delta R^*/R$ after $\tau_{\rm exp}$= 50\,min measured on the plasmonic crystal (PC) and on bare (Cd,Mg)Te. In case of the PC the data are shown for $T=10$\,K and 300\,K. As indicated, the signal amplitude on bare (Cd,Mg)Te was multiplied by a factor of 10 before plotting. The green lines are the fits of experimental data with a damped harmonic oscillation [see Eq.~(\ref{eq:fit})]. Individual curves are offset by $5\times 10^{-4}$ for better visibility. Dependence of amplitude $A$ and frequency $f$ on exposure time $\tau_{\rm exp}$ in the plasmonic crystal both at room temperature (red dots) and at 10\,K (blue squares) (b, c) and bare semiconductor at 300\,K (d, e), respectively.}
	\label{fig4}
\end{figure*}

In order to elucidate the nature of the oscillatory signal, we first perform local regression smoothing which follows the evolution of the non-oscillatory background and subtract it from the dataset. Subsequently, the remaining oscillatory part $\Delta R^*/R$, is fitted with a damped harmonic oscillation
\begin{align}\label{eq:fit}
	\Delta R^*(t)/R=A\exp\left(-\frac{t}{\tau}\right)\sin\left(2\pi f t + \phi\right)\, ,
\end{align}
where $A$ is the amplitude, $f$ is the oscillation frequency, $\tau$ is the decay time and $\phi$ is the phase. Exemplary curves are shown in Fig. \ref{fig4}(a). The phase and decay time do not depend on the exposure time $\tau_{\rm exp}$ within the accuracy of the experiment. The dependence of amplitude and oscillation frequency for the plasmonic crystal (PC) are shown in Fig.~\ref{fig4}(b) and \ref{fig4}(c), respectively. It follows that $A$ increases with $\tau_{\rm exp}$, reaching a saturation value of about $3\times 10^{-4}$ after 20\,min. The oscillation frequency $f$ experiences a decrease of about 0.1\,THz from an initial value of 3.7\,THz (3.8\,THz) at 300\,K (10K) during the first 10\,min of exposure. These dependencies follow similar behavior at both low $T=10$\,K and room temperatures. The decay time $\tau$ increases from 1.1\,ps to 2\,ps with decrease of the temperature from 300\,K to 10~K. In case of the bare semiconductor layer, the value of $A$ is about 10 times smaller at the initial stage ($\tau_{\rm exp}\approx 10$\,min) and grows only weakly without reaching the saturation value even after 120\,min of exposure [see Fig.~\ref{fig4}(d)]. The dependence of $f$ on $\tau_{\rm exp}$ is also weak as it is shown in Fig.~\ref{fig4}(e).

The THz range of the oscillations and the weak dependence of its decay time and frequency on the temperature indicate that coherent optical phonons in the semiconductor layer could be responsible for the observed transients~\cite{Zeiger92, Merlin97, Dekorsy-book}. However, none of the observed frequencies fit to the optical phonon energies at the $\Gamma$ point of the Brillouin zone in Cd$_{0.86}$Mg$_{0.14}$Te. The optical phonon in this material which is the closest in energy to the measured values is the 4.2\,THz (17.4\,meV) CdTe-like TO mode~\cite{EunsoonOh1993, Nakashima1973}. From the experiment it follows that for long exposure times the oscillation frequency approaches $f_{\rm 300K}=3.60$\,THz (14.9\,meV) at $T=300$\,K and $f_{\rm 10K}=3.72$\,THz (15.4\,meV) at $T=10$\,K which corresponds very well to the frequency of the symmetric ${\rm A}_1$ breathing mode of Te with $f_{\rm Te, 295K}=3.61$\,THz and $f_{\rm Te, 4K}=3.73$\,THz~\cite{phonons_Tellurium1998}. Indeed, formation of tellurium under optical illumination is possible in several II-VI telluride semiconductors. It has been observed in CdTe~\cite{Soares2004}, CdZnTe~\cite{Hawkins2008,Teague2009} and ZnTe~\cite{Larramendi2010,Wiedemeier2011} crystals under continuous wave illumination above the band gap. Recently, excitation of coherent phonons in a segregated layer of Te on ZnTe was induced by illuminating the semiconductor crystal with intense 10\,fs optical pulses~\cite{Shimada2014}. Thus, the buildup of the oscillating signal with increase of the pump exposure time in Fig.~\ref{fig4}(b) clearly demonstrates a change of the interface composition and formation of segregated Te increasing in amount with $\tau_{\rm exp}$. We emphasize that the long-term pump induced changes in the reflectivity transients are irreversible, i.e. the oscillatory signal in the pump-probe transients does not disappear after the intense illumination has been switched off.

\begin{figure}
	\centering
	\includegraphics[width=0.9\linewidth]{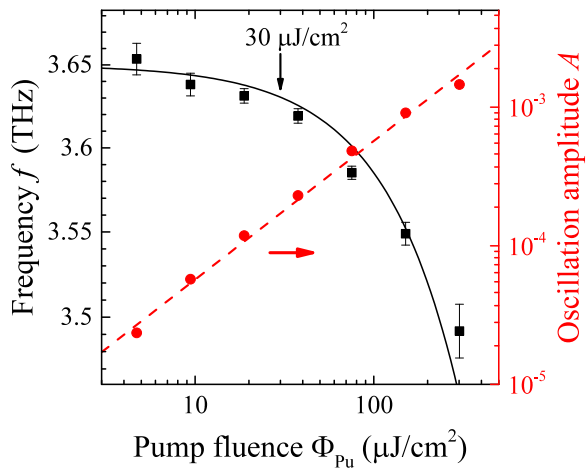}
	\caption{Dependence of the oscillation frequency $f$ and oscillation amplitude $A$ on the applied pump fluence $\Phi_{\rm pu}$ at 300\,K on the PC sample. The pump fluence used to record the transients in Figs.~\ref{fig3} and \ref{fig4} is indicated with an arrow. Softening of the ${\rm A}_1$ mode frequency follows $f=f_0 - \beta \Phi_{\rm pu}$ with $f_0=3.65$\,THz and $\beta=0.65\,{\rm cm^2 THz/mJ}$. Dashed line is a fit with a linear function $A=\alpha\Phi_{\rm pu}$, $\alpha=\text{const.}$}
	\label{fig5}
\end{figure}

After elaborating the origin of the THz oscillations, we discuss the role of SPPs. Two main features follow from the data in Fig.~\ref{fig4}. Illumination with intense laser pulses leads to (i) decrease of the oscillation frequency and (ii) saturation of the amplitude, which takes place only in the case of the PC.

At first glance, the shift of the oscillation frequency with respect to the bulk value by $\Delta f \approx 0.1$\,THz could be attributed to mode softening, i.e. electronic weakening of the crystal lattice with increase of photoexcited carrier density in Te~\cite{Hunsche1995,Hunsche1996,Misochko05}. This could occur at high excitation fluences due to stronger absorption taking place when the Te layer thickness grows. In the presence of mode softening, the oscillation frequency of the transients under prolonged pulsed illumination would be expected to be lower than the literature values $f_{\rm Te}$, which were obtained under relatively low power CW illumination. However, we observe that for long exposure times the measured oscillation frequency is in line with the literature values of the ${\rm A}_1$ phonon. In the case of the plasmonic crystal, a significant decrease of the frequency below 3.6\,THz takes place only for pump fluences above $100\,{\rm \mu J/cm^2}$ (see Fig.~\ref{fig5}) which is larger than $\Phi_{\rm pu} = 30\,{\rm \mu J/cm^2}$ used for the current experiment. Therefore, we conclude that mode softening cannot explain the observed frequency change $\Delta f$.

The elevated value for the oscillation frequency during the first 10\,min of exposure can thus be explained by a confinement of vibrational modes in a thin Te layer with thickness $w$~\cite{Sood1985}. The change of the phonon frequency $\Delta f(w)$ can be estimated as $\Delta f = f''(0)\times (\pi/w)^2/2$, where $f''(0)$ is the second derivative of the Te phonon frequency with respect to the wavevector, calculated at the $\Gamma$-point of the Brillouin zone. Naturally, $f''(0)$ depends on the direction of the layer normal, which we cannot determine. In the following estimate we adopt the value $f''(0) \approx 3\times 10^{-20}\,{\rm THz\,m^2}$ taken from fitting of the dispersion curves given by Pine and Dresselhaus~\cite{Pine1971}. Note, that $f''(0)$ is positive and, therefore, the frequency of the symmetric ${\rm A}_1$ breathing mode becomes smaller in Te layers of growing thickness $w$, which is in full accord with the experimental data.

The effect of phonon confinement allows estimating the initial thickness of the Te layer and monitoring its formation. The frequency shift at the initial stage before exposure to intensive light ($\tau_{\rm exp}=0$) corresponds to $\Delta f_{300K} = 0.12$\,THz. From this value we deduce an initial layer thickness $w \approx 1$\,nm.
Next, we can assume that the signal amplitude $A$ during the segregation process is directly proportional to the Te layer thickness.
This assumption is justified since the absorption length in Te is significantly larger ($\sim 50$\,nm)~\cite{Hunsche1995}.
The Te thickness at the initial stage corresponds to about 1\,nm with $A$ being about 30\% of the saturated amplitude value [see Fig.~\ref{fig4}(b)]. Therefore, we conclude that the formation of the Te layer stops at $w \approx 3$\,nm.

In the case of the bare (Cd,Mg)Te sample the formation of Te is less efficient and develops very slow. Here, the frequency shift remains large even after two hours of exposure $\Delta f = 0.1-0.2$\,THz corresponding to $w \approx 1-0.7$\,nm and no saturation of the amplitude is observed [see Fig.\ref{fig4}(d)]. These results directly demonstrate that the plasmonic crystal leads to a significant enhancement of the Te segregation due to SPPs.

Finally, we analyze the absolute magnitude of the oscillatory signal for long exposure times ($\tau_{\rm exp} \approx 60$\,min) which corresponds to $A = 2\times 10^{-5}$ on the bare (Cd,Mg)Te sample and $A = 3\times 10^{-4}$ in the case of the plasmonic crystal for $\Phi_{\rm pu}=30\,{\rm \mu J/cm^2}$. The oscillation amplitude $A$ evolves linearly with the pump fluence in this regime (see Fig.~\ref{fig5}). When scaled up to the pump fluences used by Hunsche {\it et al.} for the pump-probe measurements on crystalline Te~\cite{Hunsche1995,Hunsche1996} and by Kamaraju {\it et al.} on polycrystalline material~\cite{Kamaraju2010} in the same wavelength range, we find that the magnitude of $A$ in the studied hybrid Te-plasmonic crystal structures is comparable to the coherent phonon amplitudes measured in bulk tellurium. Taking into account that in our structures the layer thickness is at least one order of magnitude smaller than the absorption length of Te, we conclude that comparable signals are resulting from a smaller Te volume when SPPs are involved. This is further supported by the fact that the amplitude in the bare (Cd,Mg)Te is about one order of magnitude smaller as compared to the bulk Te data. Thus, the use of plasmonic crystals allows us to increase the sensitivity and to achieve large amplitudes of THz modulation in thin Te layers with a thickness of only several nm. The underlying mechanism of the enhancement is discussed in the following section.

\section{SPP enhancement of the differential reflectivity}
\label{sec4}

The spectral position and shape of SPP resonances in the reflectivity spectra are very sensitive to changes of the dielectric constant of the constituent materials at the interface. Therefore, differential reflection of light may be significantly enhanced in the vicinity of SPP resonances~\cite{Pacifici07, MacDonald, Brongersma, Caspers10, Pohl12}.
The role of plasmon resonances in the optical response mediated by coherent phonon oscillations has been recently studied using femtosecond pump-probe techniques.
The enhancement of coherent phonon dynamics has been demonstrated in graphite~\cite{Katayama2011} and CuI~\cite{Isshiki13} films due to local field enhancement induced by localized surface plasmons in gold nanoparticles and films with nanoscale roughness, respectively. However, for propagating SPPs in 30\,nm thin bismuth films, which were excited in Kretschmann configuration, the oscillatory signal was suppressed close to resonance conditions~\cite{Chen12}.

A fundamentally different situation is realized in our hybrid structures. Here, propagating SPPs originate from the interface between a noble metal (Au) and a transparent semiconductor (Cd,Mg)Te with an only few nm thick Te layer at the interface between them. Therefore, optical losses in our hybrid system are expected to be significantly lower. In order to check the role of SPPs on the strength of the oscillatory signal, we perform two types of experiments. While the first is based on the polarization selectivity of SPPs in plasmonic crystals consisting of linear gratings, the second relies on the spectral dependence of the differential reflectivity close to SPP resonances. Both experiments indicate that SPPs lead to an enhancement of the oscillatory signal by at least one order of magnitude. All measurements were performed at $\Phi_{\rm pu} = 140\mu$J/cm$^2$ after waiting about 60\,min until the formation of the Te layer was fully accomplished.

\begin{figure}
	\centering
	\includegraphics[width=0.7\linewidth]{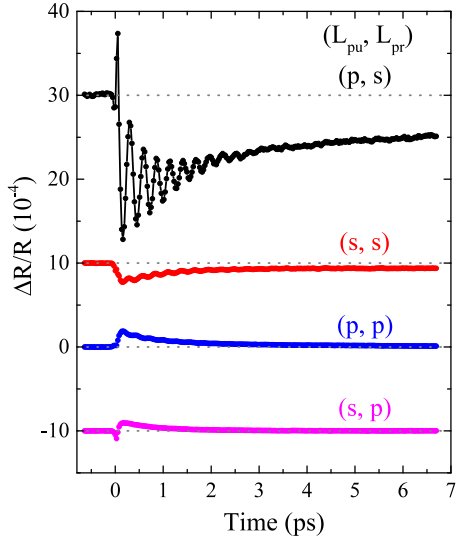}
	\caption{Differential reflectivity transients measured for various pump and probe polarization configurations at $T=300$\,K. The orientation of the plasmonic crystal with respect to the incident beams corresponds to $\varphi_{\rm pu} = 0^\circ$ and $\varphi_{\rm pr} = 90^\circ$ as shown in Fig.~\ref{fig2}(a). The linear $p-$ or $s-$polarizations of the pump $L_{\rm pu}$ and the probe $L_{\rm pr}$ beams are labeled by ($L_{\rm pu},L_{\rm pr}$). Measurements were performed after an exposure time of 60\,min at a pump fluence of $\Phi_{\rm pu} = 140\,{\rm \mu J/cm^2}$ when the oscillatory signal was fully built up. The transients are offset for better visibility. The dotted lines indicate the zero levels of the individual curves.}
	\label{fig6}
\end{figure}

Use of plasmonic crystals based on a linear grating allows selecting a proper direction of linearly polarized pump and probe beams in order to compare the transient signals with and without excitation of SPPs with the corresponding beams. The set of $\Delta R(t)/R$ transients in the hybrid plasmonic crystal after formation of the Te layer for $\varphi_{\rm pu} = 0^\circ$ and $\varphi_{\rm pr} = 90^\circ$ is shown in Fig.~\ref{fig6}.
The magnitude of the oscillatory signal in the hybrid plasmonic crystal is very sensitive to the polarizations $L_{\rm pu}$ and $L_{\rm pr}$ of both beams. The orientation of the plasmonic crystal with respect to the incident beams [see Fig.~\ref{fig2}(a)] dictates that SPPs are excited when the pump is $p$-polarized and the probe is $s$-polarized, i.e. ($p,s$) configuration. The strongest modulation of the differential reflectivity is observed exactly in this configuration. When either one of the beams or both of them are switched to another non-plasmonic polarization, the THz-oscillations practically vanish. In contrast to that, the differential reflection signal of unpatterned (Cd,Mg)Te is about the same in all four polarization configurations. This observation proves the crucial role of SPPs for the excitation of coherent phonons in the tellurium layer, as well as for their efficient detection.

\begin{figure}
	\centering
	\includegraphics[width=0.9\linewidth]{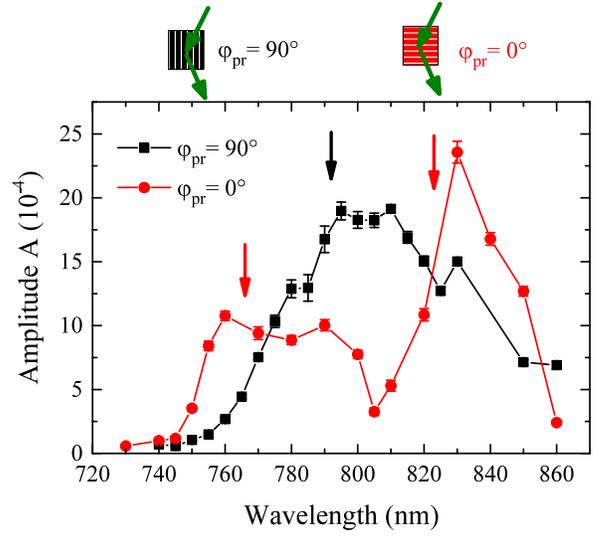}
	\caption{Amplitude of the oscillatory signal resolved by wavelength for two different sample orientations corresponding to excitation of SPPs by the probe beam with $\varphi_{\rm pr}=90^\circ$ (black) and for $\varphi_{\rm pr}=0^\circ$ (red). $\Phi_{\rm pu} = 140\,{\rm \mu J/cm ^2}$. Spectral positions of the calculated SPP resonances are indicated with arrows.}
	\label{fig7}
\end{figure}

\begin{figure*}
	\centering
	\includegraphics[width=0.8\linewidth]{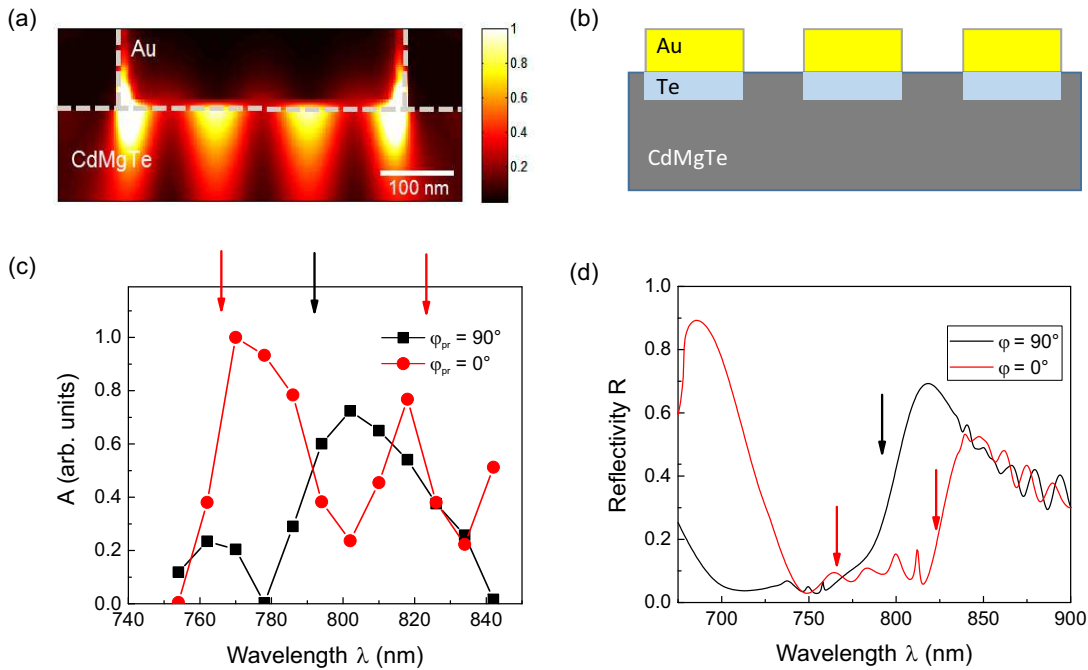}
	\caption{Results of the electromagnetic modeling. (a)  The distribution of the electric part of the electromagnetic energy density in the plasmonic crystal and specifically in the (Cd,Mg)Te part before the Te segregation for illumination with laser light having a Gaussian spectral profile of full width at half maximum $\Delta \lambda=63$\,nm centered at $\lambda=800$\,nm and incident under $\theta=17^\circ$, $\varphi=90^\circ$. (b) Schematic of the plasmonic crystals after Te formation. (c) Spectra of the oscillatory signal amplitude for two different sample orientations corresponding to excitation of SPPs by the probe beam with $\varphi_{\rm pr}=90^\circ$ (black) and for $\varphi_{\rm pr}=0^\circ$ (red). (d) White light reflectivity spectra calculated for ($\theta=17^\circ$, $\varphi=90^\circ$) and ($\theta=17^\circ$, $\varphi=0^\circ$) via the RCWA method. Spectral positions of the SPP resonances in (c) and (d) are indicated with arrows.}
	\label{fig8}
\end{figure*}

Another approach, which allows determining the role of SPPs in detection, is based on the analysis of the spectral dependence of differential reflectivity transients. Here, we expect that an enhancement of the probe beam modulation should take place in the vicinity of the SPP resonances. Spectrally resolved measurements were performed using a tunable liquid crystal based interference filter with a bandwidth of 7\,nm in the reflected beam path. Two different excitation geometries were studied by orienting the grating slits of the plasmonic crystal either parallel ($\varphi_{\rm pr}=90^{\circ}$) or perpendicular ($\varphi_{\rm pr}=0^{\circ}$) to the incidence plane of the probe beam (see Fig.~\ref{fig2}). The plasmonic polarization configurations in these two geometries were $(p,s)$ and $(s,p)$, respectively. Each of the geometries addresses different SPP resonances in detection. The amplitude $A$ was extracted from the transients using the same procedure as described in Sec.~\ref{sec3} using Eq.~(\ref{eq:fit}) and the results are shown in Fig.~\ref{fig7}. It is clearly seen that $A(\lambda)$ differs strongly for the two geometries. While for $\varphi_{\rm pr}=90^\circ$ the maximum enhancement is observed at a wavelength of about 800\,nm, the spectral dependence at $\varphi_{\rm pr}=0^\circ$ shows a local minimum for that wavelength and enhanced amplitudes at 830\,nm and 760\,nm.

A theoretical description of the spectral dependence of the transient signal was performed using the Rigorous Coupled-Wave Analysis (RCWA)~\cite{Li03}. For this purpose, we first calculate the distribution of the electric part of the electromagnetic energy density in the plasmonic crystal and specifically in the (Cd,Mg)Te part before Te segregation takes place which is shown in Fig.~\ref{fig8}(a). It follows, that the electromagnetic energy is mostly concentrated under the gold stripes. Therefore, for the calculations we assume that formation of Te takes place in those regions [Fig.~\ref{fig8}(b)].

In the perturbation regime the differential reflectivity of the oscillatory part can be written as
\begin{equation}
	\frac{\Delta R^*}{R} = \left[\frac { C_1(\lambda)}{\epsilon_1}\frac{d\epsilon_1}{dQ} + \frac{C_2(\lambda)}{\epsilon_2}\frac{d\epsilon_2}{dQ} \right]\delta Q\, ,
	\label{eq:dR}
\end{equation}
where $\delta Q(t)=a\exp(-t/\tau)\sin(2 \pi ft + \phi)$ is the time-dependent displacement of the ${\rm A}_1$ nuclei coordinate, which corresponds to a coherent oscillation of the ${\rm A}_1$ phonon. Here, $\epsilon_1 = 31$ and $\epsilon_2 = 19$, are the unperturbed real and imaginary parts of the dielectric constant in the spectral region of interest, respectively ~\cite{Palik}.
For the overall shape of the spectral dependence, the ratio between the modulation of the real part $d\epsilon_1/dQ$ and the modulation of the imaginary part $d\epsilon_2/dQ$ by the coherent phonon motion is of prime importance. For the modeling presented in Fig.~\ref{fig8}(c) the relation $d\epsilon_2/dQ\approx-1.5(d\epsilon_1/dQ)$
was used, which was taken from the data in bulk tellurium~\cite{Mazur03,Roeser04}.  	
The spectral dependencies of $C_1(\lambda)$ and $C_2(\lambda)$ are determined by the parameters of the plasmonic crystal. Here, the thickness $w$ of the segregated Te layer enters due to its influence on the complex index of refraction at the gold-semiconductor interface. Calculation of $C_1(\lambda)$ and $C_2(\lambda)$ with the use of Eq.~(\ref{eq:dR}) allows modeling the spectrum of the oscillation amplitude [see Fig.~\ref{fig8}(c)]. It agrees qualitatively with the experimental data in  Fig.~\ref{fig7} if the thickness of the segregated Te regions is $w=3.5$\,nm, which is in excellent agreement to the above estimations based on the oscillation frequency.

When the plane of incidence of the probe beam is parallel to the slits of the plasmonic crystal ($\varphi_{\rm pr}=90^\circ$) both experimental and calculated spectra of the oscillation amplitude have a maximum at around 795\,nm,  which is close to the SPP resonance at the semiconductor interface ($\lambda = 792$\,nm). When the probe beam is incident perpendicularly to the slits ($\varphi_{\rm pr}=0^\circ$) the amplitude spectra have two peaks. Their spectral positions in the experimentally obtained and calculated data are slightly different. Nevertheless, in both cases the peaks are in close proximity to the SPP resonances at around 823\,nm and 766\,nm. This finding additionally supports the assumption that plasmonic enhancement is responsible for the increased differential reflectivity modulations in the hybrid structure. The observed amplitude peaks are close to the extrema of the reflectivity spectrum [see Fig.~\ref{fig8}d)] rather than to its slopes. Consequently, the enhancement is mainly caused by a broadening of the SPP resonance due to the variation of the Te dielectric permittivity. Changes in the dielectric permittivity might also cause the SPP resonance to shift. However, in the case of the highly absorptive Te regions this contribution to the differential reflectivity is much less pronounced.

\section{Conclusions}
\label{sec5}
We demonstrate that the presence of propagating surface plasmon polaritons at the interface between a noble metal grating and a II-VI semiconductor enhances the tellurium segregation under sub-bandgap illumination. The formation of elemental tellurium takes place under the metal stripes and saturates at a thickness of about 3\,nm for resonant excitation of surface plasmon polaritons after about 20 minutes for an illumination with $140\,{\rm \mu J/cm^2}$.

Next, a significant enhancement of the excitation and detection of coherent optical phonons observed by pump-probe transients is manifested in the THz modulation of the SPP spectral width (quality factor). The overall sensitivity to lattice vibrations is increased by more than one order of magnitude. This allows measuring the coherent phonon dynamics in a very thin Te layer (1\,nm) and resolving the confinement of the ${\rm A}_1$ symmetry optical phonon mode leading to an increase of the phonon frequency by $0.1-0.2\,$THz. The modulation of surface plasmon resonances at THz frequencies with the help of optical phonons holds promise for application in ultrafast active plasmonics. In this respect, the use of optical phonons is appealing because the amplitude of lattice motion can be coherently controlled by a train of optical pulses.

\begin{acknowledgments}
	The work was supported by the Deutsche Forschungsgemeinschaft (Project No. AK40/7-1), by the German Ministry of Education and Research (BMBF) (FKZ: 05K13PE1) and by the Russian Foundation for Basic Research (13-02-91334). The project ‘SPANGL4Q’ acknowledges financial support from the Future and Emerging Technologies (FET) programme within the Seventh Framework Programme for Research of the European Commission, under FET-Open grant no. FP7-284743. The research in Poland was partially supported by the National Science Centre (Poland) through Grants No. DEC-2012/06/A/ST3/00247 and No.DEC-2014/14/M/ST3/00484. V.I.B. acknowledges support from the Alexander von Humboldt Foundation.
	\end{acknowledgments}

\end{document}